\documentclass[aps,prd,showpacs,amsfonts]{revtex4-1}
\usepackage{epsfig}
\usepackage{graphics}
\usepackage{amsmath}
\newcommand{\Eq}[1]{Eq.~(\ref{#1})}
\newcommand{\Eqs}[1]{Eqs.~(\ref{#1})}

\newcommand{\be}{\begin{equation}}
\newcommand{\ee}{\end{equation}}
\newcommand{\bea}{\begin{eqnarray}}
\newcommand{\eea}{\end{eqnarray}}

\begin{document}

\author{Jacob D. Bekenstein}
\affiliation{Racah Institute of Physics, Hebrew University of
Jerusalem, Jerusalem 91904, Israel\\}\date{\today}
\title{Statistics of black hole radiance and the horizon area spectrum}
\begin{abstract}
 The statistical response of a Kerr black hole to incoming quantum radiation has heretofore been studied by the methods of maximum entropy or quantum field theory in curved space-time.  Neither approach pretends to take into account the quantum structure of the black hole itself.  To address this last issue we calculate here the conditional probability distribution associated with the hole's response by assuming that the horizon area has a discrete quantum spectrum, and that its quantum evolution corresponds to jumps between adjacent area eigenvalues, possibly occurring in series, with consequent emission or absorption of quanta, possibly in the same mode.  This ``atomic model'' of the black hole is implemented in two different ways and recovers the previously calculated radiation statistics in both cases.  The corresponding conditional probably distribution is here expressed in closed form in terms of an hypergeometric function.

\end{abstract}
\maketitle

\section{Introduction}
\label{sec:intro}

The Hawking radiance of a black hole is usually characterized by the mean number of quanta emitted in each mode~\cite{Hawking0}.  A more precise characterization entails specification of the probability distribution of the number of quanta in each mode.  Such distributions were first calculated early~\cite{Parker,Wald,Hawking}.  In later years the issue of unitary evolution has convinced many workers that the radiance must be in a pure quantum state.  Full characterization of such state would entail, at the very least, giving the \emph{joint} probability distribution of the number of quanta emitted in the various modes.  Since the purity of the radiance raises still unsolved problems, we focus here on the one-mode marginal probability distributions; by marginalizing the joint distribution we remain neutral with regard to the ultimate outcome of the said controversy.  

More specifically, we here investigate the relation between the one-mode response of a Kerr black hole to incident bosonic radiation and the energy spectrum for Kerr black holes.  The question we focus on is whether any insight into quantum gravity can be had from such investigation. The ready answer would be that the thermality of the Hawking radiance, which seems incontestable if the purity status is rejected, should make it blind to any specifics of the black hole structure and dynamics, so that nothing can be learned from it about quantum gravity.  In our view this is a simplistic response.  To draw an analogy, the radiation emitted by an atom says a lot about its quantum structure. Only when numerous atoms radiate together incoherently does the radiation assume a thermal guise. Now to some extent the black hole, in its simplicity, is the hydrogen atom of quantum gravity~\cite{BekMG,Corda}.  By analogy we could expect to glean insights into this subject from detailed analysis of the black hole radiance which, after all, emanates from a coherent gravitational structure, not from an assembly of quantum systems.  Indeed, as we shall show, a definite assumption about the horizon area spectrum of a black hole recovers, by simple argumentation, the correct one-mode response probability distribution.

We shall focus here on the model of a quantum black hole espoused in Refs.~\cite{spectrum,Mukh,BekMukh}.   Its central claim is that the horizon area $A$ (or entropy $S$) is quantized in equally spaced steps.  For the Kerr black hole (mass $M$ and angular momentum $J$) this means (we take $c=1$)
\be
A=4\pi G^2 \big(M+\sqrt{M^2-J^2/G}\big)^2 =\alpha L_P{}^2\, n;\quad  n=1,2,\cdots,
\label{Aq}
\ee
where $\alpha$ is a dimensionless positive constant.  Two possible values of $\alpha$ are often mentioned: $\alpha=8\pi$~\cite{spectrum,Maggiore} and $\alpha=4 \log k$ with $k=2,3,\cdots$~\cite{Mukh,BekMukh,Hod}.  We shall here focus on macroscopic holes, those with very large $n$. We assume that the fundamental transition of a black hole with consequent emission or absorption of one radiation quantum is one with $n\Rightarrow n\mp1$.  Conservation laws then fix the frequency and azimuthal quantum number of the quanta; for emission $\hbar\omega=-\Delta M$ and $\hbar\tilde m=-\Delta J$ with $\tilde m\in (\cdots,-2,-1,0,1,2,\cdots)$.  Thus differencing (more practically differentiating) \Eq{Aq} gives as the characteristic parameter for the fundamental transition
\be
x\equiv \hbar(\omega-\tilde m\Omega)/T_{\rm H} =\alpha,
\label{x}
\ee
where $T_{\rm H}$ and $\Omega$ are the Hawking temperature and rotational angular frequency for the Kerr black hole.  [Use has been made of the area formula, $S_{\rm BH}=A(4L_P{}^2)^{-1}$, as well as the thermodynamic relations $(\partial S_{\rm BH}/\partial M)_J=1/T_{\rm H}$ and $(\partial S_{\rm BH}/\partial J)_M=-\Omega/T_{\rm H}$].  It may be seen that the fundamental transition's $x$ parameter, or $\omega$ for fixed $\tilde m$, is constant over a moderate range of $M$ or $A$.  This means that a moderately long series of quanta with like values of $\omega$ and $\tilde m$ can be emitted by a macroscopic black hole making successive $n\Rightarrow n-1$ transitions before the frequency begins to shift and a switch to another mode can occur.  Likewise, transitions with $n\Rightarrow n-2, n\Rightarrow n-3, \cdots$ can also take place, with values of the parameter $x$ which are integral multiples of that in \Eq{x}.  These too can occur in series.
Every such series gives rise to a spectral line; there are many spectral lines for $n\Rightarrow n-1$ (likewise for $n\Rightarrow n-2$, etc.), each for a different value of $\tilde m$.

We remark that any other quantization rule according to which $A$ is a smooth function of a quantum number analogous to $n$ will lead to similar conclusions regarding the allowed values of $\omega$.  Not so when a black hole parameter different from $A$ is subject to quantization. Likewise, completely different results would be expected for a continuum horizon area spectrum.

\section{Jump probabilities}
\label{sec:jump}

From now on we refer to transitions between $A$ levels as jumps.  Let us focus on two adjacent area levels of an isolated black hole, $u$ and $l$.  Each quantum state belonging to the upper level $u$ has a definite probability to jump-down to a state belonging to the lower level $l$ whereby a quantum with a definite $\tilde m$ is emitted; we denote this \emph{elementary probability} by $e^{-\beta}$ with $\beta$ real and positive~\cite{Fulfill}; to reduce clutter we do not the $x$ or $\tilde m$ of the mode.  Regarding the time span to which this probability refers, we shall take it as the time associated with emission of the mode of black hole radiation which emerges from the said transition.  Of course the time depends on how monochromatic we take the mode to be; very monochromatic modes are emitted over a long time.    Once the black hole has jumped down one area level, it is in a situation similar to the initial one, so another jump to one level lower should also take place with the same probability $e^{-\beta}$.  This jumping down can repeat with the quanta emitted from each belonging to the same mode.  However, once $A$ has changed significantly, the emission can be considered to have shifted to a different mode since $x$ will also have changed somewhat.  Similar remarks apply to jumps with $n\Rightarrow n-2, n\Rightarrow n-3,\cdots$.

It is an assumption here that the chance of a jump is uninfluenced by whether the present level was the initial one or not, the chain of jumps being regarded as a quantum Markov process.   Obviously, with probability $1-e^{-\beta}$ no jump takes place during the mode's time span, thus interrupting the chain.    Accordingly, the probability for $j$ successive jump-downs with given $x$ and $\tilde m$ is
\be
p_{\downarrow j}= (1-e^{-\beta})\,e^{-\beta j}.
\label{spontaneous} 
\ee
This probability distribution is already normalized: $\sum_{j=0}^\infty p_{\downarrow j}=1$.

The probability $e^{-\beta}$ refers to a spontaneous jump-down in the area spectrum with emission of one quantum.  There should also be an \emph{elementary probability} for the hole in some state belonging to $l$  to jump-up to any state belonging to $u$~\cite{BekSchiff} when a quantum of the appropriate $x$ and $\tilde m$ is incident.  We denote this probability by $e^{-\mu}$ with $\mu$ also real and positive.  

Now $e^{-\mu}$ and $e^{-\beta}$ are connected via the principle of detailed balance.  For example, we can consider the hole in equilibrium with a radiation bath at temperature $T_{\rm H}$.   Then according to Planck the mean number of quanta incident on the black hole in a radiation mode labeled by $x$ is $(e^x-1)^{-1}$; we, of course, exclude the customary phase-space factor-it relates to the number of modes. The mean rate at which the hole jumps from the state in $l$ to that in $u$ must be proportional to this mean number.  It will be shown in Sec.~\ref{sec:stimulated} that spontaneous emission must be accompanied by stimulated emission.  We assume that the hole's jump-down probability is uninfluenced by whether quanta in the relevant mode are already incident. Thus we can write
\be
e^{-\mu}(e^x-1)^{-1}=e^{-\beta}+e^{-\beta}(e^x-1)^{-1},
\ee
which tells us that
\be
e^{-\mu}=e^{-\beta+x}.
\label{detailed}
\ee

With $e^{-\mu}$ in hand we can now calculate the probability of total absorption of $k$ quanta incident on the black hole in a single mode whose $x$ matches the spacing between levels $u$ and $l$.  Assuming that, as long as there are quanta that can be absorbed, the probability of a jump-up of the hole is independent of what happened before, we get for the probability of $k$ jump-ups in sequence with consequent absorption of $k$ quanta
\be
p_{\uparrow k}=(1-e^{-\beta})e^{-\mu k};\qquad k=1,2,\cdots .
\label{pup}
\ee
The factor $(1-e^{-\beta})$ again represents the probability that the hole does not experience a jump-down which would interrupt the sequence of $k$ jumps-up.   Note that in the present case the sum over $k$ in \Eq{pup} need not be unity since different $k$s do not label different outcomes of one initial set up, but rather different initial set ups.

Of course an incident quanta may fail to be absorbed, i.e. the black hole may not jump-up in its presence.  Now, the probability complementary to $e^{-\mu}$ is $1-e^{-\mu}$.  This is not the full contribution; the black hole may also spontaneously jump-down while failing to jump-up.  Thus the probability that the black hole avoids either and remains in its initial state despite the presence of an incident quantum is
\be
p_0=(1-e^{-\mu})(1-e^{-\beta}).
\label{no}
\ee

\section{Statistics of absorption and emission}
\label{sec:statistics}

As already hinted, \Eq{spontaneous} can be regarded as the probability of \emph{spontaneous} emission of $j$ quanta in the one mode considered.  We could write this as $p(j|0)$, where $p(m|n)$ is  generally defined as the \emph{conditional probability} that $m$ quanta are emitted when $n$ are incident in the mode in question~\cite{BekMeis}.  Of course we must require $\sum_{m=0}^\infty p(m|n)=1$ for all $n$.  Likewise we may interpret \Eq{pup} to give $p(0|k)$, the probability that if $k$ quanta impinge on the black hole in a given mode, all are absorbed.  We now compute $p(m|n)$ for all other cases.

Suppose $n$ indistinguishable quanta are incident on the black hole in one mode.  And suppose $k\leq n$ of these fail to cause any jump-up of the black hole, and thus survive to occupy the outgoing counterpart of the incident mode. Assuming that each quantum acts independently and taking \Eq{no} into account, the probability for this partial outcome must be $(1-e^{-\mu})^k(1-e^{-\beta})^k$.    The probability that the hole makes $n-k$ consecutive jump-ups as it absorbs the other $n-k$ incident quanta is evidently $e^{-\mu (n-k)}$.   And the number of independent ways in which the $n$ incident quanta can be partitioned into absorbed and surviving quanta is $n!/[(n-k)!k!]$.  
We thus get the \emph{partial} probability (as always $0!=1$)
\be
\frac{n!}{(n-k)!k!}(1-e^{-\mu})^k(1-e^{-\beta})^k e^{-\mu (n-k)}.
\ee

The number of quanta in the  outgoing counterpart of the said mode is $m$.  Of these $k\leq m$ are surviving incident ones.  Thus the hole must emit $m-k$ quanta as it jumps down $m-k$ levels.  According to \Eq{spontaneous} the  probability for this is $(1-e^{-\beta})e^{-\beta (m-k)}$.  The number of independent ways in which the $m!$ outgoers can be selected from surviving and freshly emitted quanta is $m!/[(m-k)!k!]$.  We thus have the second partial probability
\be
\frac{m!}{(m-k)!k!}(1-e^{-\beta}) e^{-\beta (m-k)}.
\ee
The conditional emission probability $p(m|n)$ must obviously contain the product of the above partial probabilities.  In addition, the number $k$, being not observable, must be summed over from zero to the smaller of $n$ or $m$.  Thus
\bea
p(m|n)&=&(1-e^{-\beta})e^{-(\mu n+\beta m)}\sum_{k=0}^{{\rm min}(m,n)}\frac{n!m!X^k}{(n-k)!(m-k)(k!)^2};
\label{pmn}
\\
X&\equiv &(e^\beta-1)(e^\mu-1).
\label{X}
\eea

To check this result we first set $m=0$; we get $p(0,n)=(1-e^{-\beta})e^{-\mu n}$ which is what we would expect from \Eq{pup}.  Setting rather $n=0$ we get $p(m,0)=(1-e^{-\beta})e^{-\beta m}$ which is the expected result for spontaneous emission, i.e., \Eq{spontaneous}.  We next check that $p(m|n)$ is normalized.  Summing $p(m|n)$ over all $m$ and interchanging the two sums while respecting the constraints $k\leq m,n$ gives
\be
\sum_{m=0}^\infty p(m|n)=(1-e^{-\beta})e^{-\mu n}\sum_{k=0}^n\frac{n!X^k}{(n-k)!(k!)^2}\sum_{m=k}^\infty \frac{m!}{(m-k)!}e^{-\beta m}.
\label{inter}
\ee
The inner sum is done in Appendix A (\Eq{first}); substituting it here gives
\be
\sum_{m=0}^\infty p(m|n)=\sum_{k=0}^n\frac{n!}{(n-k)!k!}(1-e^{-\mu})^k e^{-\mu (n-k)},
\ee
which equals unity by Newton's binomial theorem. We remark that the sum in \Eq{pmn} is symmetric in $m$ and $n$ and can, in fact, be identified with the hypergeometric function ${}_2 F_1(-m,-n,1,X)$~\cite{AbraSteg}.  The nonpositive integer nature of the first two arguments is responsible for cutting off the hypergeometric series and rendering it a polynomial in $X$; this turns out to be the Jacobi polynomial $P^{(0,-m-n-1)}(1-2X)$ (Eq.~15.4.6 of ref.~\cite{AbraSteg}).  

We now express $e^{-\beta}$ and $e^{-\mu}$, the microscopic parameters of the black hole transition,  in terms of observable black hole parameters: $x$, as defined above, and $\Gamma$, the so called absorptivity of the black hole in the said mode (also known as the grey-body factor or barrier tunneling coefficient).  First we calculate the mean number of quanta spontaneously emitted as a result of the jumps associated with the said mode (see \Eq{spontaneous}):
\be
\langle m\rangle_{\rm sp}=\sum_{m=0}^\infty (1-e^{-\beta})\,m\,e^{-\beta m}=(e^\beta-1)^{-1}.
\label{mean0}
\ee 
In Hawking radiance this mean is $\Gamma$ times what could be attributed to a \emph{blackbody} with temperature $T=T_{\rm H}$.  Although in the present paper the black hole spectrum is regarded as being made up of (broadened) lines, we can still use \Eq{mean0} for the modes making up each line. 

Now an ideal black body would emit into the mode a Planck's mean number,
\be
\langle m\rangle_{\rm bb}=(e^x-1)^{-1},
\ee
of quanta.  Thus the black hole spontaneously emits a mean number
\be
\langle m\rangle_{\rm sp}=\Gamma(e^x-1)^{-1}.
\label{spon}
\ee
Comparing this with \Eq{mean0} gives
\be
e^{-\beta}=[1+(e^x-1)\Gamma^{-1}]^{-1}.
\label{embeta1}
\ee

Obviously for emission from a Schwarzschild black hole $x>0$ and $\Gamma<1$; hence $e^{-\beta}$ turns out to be smaller than unity, as it should.  For emission from the Kerr black hole $x=\hbar(\omega-\tilde m\Omega)/T_{\rm H}$.  For modes with $x>0$ the story is the same as for the Schwarzschild case.  Modes with $x<0$ are known as superradiant modes because incident radiation in such modes is amplified by the hole, so that $\Gamma<0$ for every such mode.  For these modes it again follows from \Eq{embeta1} that $e^{-\beta}<1$.

Substitution of \Eq{embeta1} and (\ref{detailed}) into \Eq{pmn} now gives us
\be
p(m|n)=\frac{(e^x-1)e^{xn}\Gamma^{m+n}}{(e^x-1+\Gamma)^{m+n+1}}\ {}_2 F_1(-m,-n,1,X).
\label{pmn1}
\ee
This expression for the conditional probability $p(m|n)$ is our central result here.  To recapitulate, it describes the statistics of black hole radiance in the particular mode characterized by the values of $x$ and $\Gamma$.  It results from a set of simple assumptions as to the quantum jumps that the black hole can undergo, either spontaneous jumps, or those made possible by incident quanta in the same mode. 

\section{Comparison with the early results}
\label{sec:comparison}

The consistency of that set of assumptions will now be verified by comparison of \Eq{pmn1} with the expression for the conditional probability obtained by two independent methods that do not focus on the black hole's quantum structure. The $p(m|n)$ was first calculated in Ref.~\cite{BekMeis} by applying the maximum entropy method of information theory to the statistics of radiation outgoing from a black hole which is bathed by external thermal radiation.  In Ref.~\cite{PananWald} the same result was obtained by applying methods of quantum field theory in curved spacetime to the radiance.  Both of these approaches focus on the radiation, and treat the black hole as a fixed background.  It may be noted that an ideal material grey body also exhibits the same conditional probabilities~\cite{BekSchiff}.

The results of Refs.~\cite{BekMeis} and~\cite{PananWald} can both be put in the form
\bea
p(m|n)&=&\frac{(e^x-1)e^{xn}\Gamma^{m+n}}{(e^x-1+\Gamma)^{m+n+1}}\sum_{k=0}^{{\rm min}(m,n)}\frac{(m+n-k)!(X-1)^k}{(n-k)!(m-k)!k!};
\label{pmn2}
\\
X&=&\frac{(e^x-1)^2e^{-x}(1-\Gamma)}{\Gamma^2},
\label{X2}
\eea
where we have denoted the parameter also by $X$ since by \Eqs{detailed} and (\ref{embeta1}) it is equivalent to that in \Eq{X}.
Now the sum in \Eq{pmn2} may be identified with 
\be
\frac{(m+n)!}{m!n!}{}_2F_1(-m,-n,-m-n,X-1);
\label{F}
\ee
this is again a polynomial because the first two arguments are nonpositive integers.  The expression in \Eq{F} is identical to the hypergeometric function appearing in \Eq{pmn1}.
Hence, the expressions for $p(m|n)$ in \Eqs{pmn1} and (\ref{pmn2}) are identical.

This agreement shows that the assumptions made in Secs.~\ref{sec:jump} and \ref{sec:statistics} regarding the black hole area spectrum and transition dynamics are consistent with what is already known about black hole radiance, regardless of whether the hole is in vacuum or immersed in radiation.  In the previous section we composed $p(m|n)$ from the probability distributions for quanta which miss being absorbed in a black hole jump-up and for quanta which are emitted as a result of a hole jump-down.  In the next sections we shall rather compose $p(m|n)$ from the probability distributions for scattering, spontaneous and stimulated emission of quanta; this last approach was introduced in Ref.~\cite{Fulfill}.

\section{Scattering}
\label{sec:scattering}

Imagine an incident quantum in a definite mode.  With probability $e^{-\mu}$ it gets absorbed concurrently with a jump-up of the black hole.  This may be followed by a jump-down of the black hole or several such in sequence.  The total probability of these independent events is
\be
\Gamma_0=e^{-\mu}(1+e^{-\beta}+e^{-2\beta}+\cdots)=e^{-\mu}(1-e^{-\beta})^{-1}.
\label{Gamma0}
\ee
Evidently, this is the probability that the incident quantum disappears into the hole with all possible consequences for the hole.  We denote it by $\Gamma_0$ because in a real sense it is the true absorption probability for a quantum of the given mode whereas, as we shall show presently, $\Gamma$ stands for a somewhat different thing.  

An external observer unaware of the inner workings of the black hole is surely aware of scattering off the hole's geometry.  What is the scattering probability distribution?  Following the above tack, we argue that since $\Gamma_0$ is the absorption probability, $1-\Gamma_0$ must be the probability that a quantum incident on the black hole in the given mode is scattered (reflected) into the outgoing form of the mode  regardless of what happened to the hole. Conservation of energy and angular momentum insure that the quantum remains associated with the same mode. Now if $n$ indistinguishable quanta are incident in one mode, the probability that a number $l\leq n$ of these are scattered  is evidently
\be
p_{\rm sc}(l|n)=\frac{n!}{l! (n-l)!}(1-\Gamma_0)^l\, \Gamma_0{}^{n-l}.
\label{scatt}
\ee
Use of \Eq{Gamma0} allows us to put the desired distribution in terms of observable parameters:
\be
p_{\rm sc}(l|n)=\frac{n!}{l! (n-l)!} \frac{e^{-\mu n}e^{-\beta l}(X-1)^{l} }{(1-e^{-\beta})^n}.
\ee

\section{Stimulated emission}
\label{sec:stimulated}

If we replace $e^{-\mu}$ in \Eq{Gamma0} by means of \Eq{detailed} and then replace $e^{-\beta}$ in the result with help of \Eq{embeta1} we get
\be
\Gamma=\Gamma_0(1-e^{-x})
\label{Gamma1}.
\ee
This result already shows that a black hole is capable of stimulated emission~\cite{BekMeis,Fulfill}.  To see this consider how $\Gamma$ is calculated in practice.  One imagines a classical wave in the form of a mode function of the appropriate field directed onto the hole; $1-\Gamma$ is identified with the fraction of the energy of the initial wave which is returned outward as a result of the scattering by the hole.   Recall that $\Gamma_0$, being a probability, is always positive.  Now for a mode with $x<0$, \Eq{Gamma1} predicts that $\Gamma<0$ so that $1-\Gamma>1$: the incident wave is thus predicted to be amplified, a sure sign of stimulated emission.  The corresponding modes are called super radiant, and are well studied numerically.

Turn now to a mode with $x>0$. \Eq{Gamma1} shows that $0<\Gamma<1$ so that $1-\Gamma$ \emph{exceeds} the scattering probability $1-\Gamma_0$ (the ratio of the two last two being independent of the wave's initial amplitude).  How is this possible?  Only if the black hole is stimulated by the incident wave to emit the same kind of wave and thus strengthen the outgoing wave in direct proportion to the incident wave's strength.  In this non-superradiant mode  the outgoing wave is weaker than the incident one, but not as weak as would be expected from the size of $\Gamma_0$.  We must conclude~\cite{BekMeis,Fulfill} that stimulated emission occurs in these modes also, but with strength insufficient to actually amplify the incident wave.  

Turning to our main concern, we ask what is $p_{\rm st}(m|n)$, the conditional probability for emission of $m$ quanta given that $n$ are incident in the same mode?  In ref.~\cite{Fulfill} it was obtained by judiciously decomposing $p(m|n)$ into a the scattering contribution from \Eq{scatt} and a part which could be interpreted in terms of emission.  Here we shall derive  $p_{\rm st}(m|n)$ using an approach like the one we followed to obtain $p(m|n)$ in Sec.~\ref{sec:statistics} of the present paper.

Stimulated emission means that each of the $n$ incident quanta may be responsible for emission of some of the outgoing $m$ quanta.  In how many ways can one associate the outgoing quanta, each to some incident one, all this in harmony with the bosonic character of the quanta?  Recall the standard textbook question, in how many ways can one arrange $m$ bosons in $g$ cells given that the bosons are identical?~\cite{LLSP}.  The answer is 
\be
\frac{(m+g-1)!}{m!(g-1)!}.
\label{comb}
\ee  
This last problem maps directly into ours; we get the desired number by replacing $g\Rightarrow n$ above.

We shall assume that the dynamics of black-hole jump-down and quantum emission is the same whether occurring spontaneously or whether it is induced by an incident quantum.  This is in keeping with what we know from atomic physics (equality of the properly defined Einstein coefficients for spontaneous and stimulated emission).  The one difference here is that it is possible for the black hole to be induced to make several jump-downs in sequence and place the consequently emitted quanta in the same mode.  (This last is very rare, if at all possible, in atomic physics.)  To the $m_j$ emitted quanta associated with incident quantum $j$ we must associate a probability $e^{-\beta m_j}$ corresponding to $m_j$ successive black hole jump-downs.  After those the jumping-down is arrested with probability $1-e^{-\beta}$.  Thus the overall probability $p_{\rm st}(m|n)$ is given by
\be
(1-e^{-\beta})e^{-\beta m_1}\cdot (1-e^{-\beta})e^{-\beta m_2} \cdots (1-e^{-\beta})e^{-\beta m_{n}}.
\ee
In light of \Eq{comb} and the fact that $\sum_j m_j=m$, we have our final formula
\be
p_{\rm st}(m|n)=\frac{(m+n-1)!}{m!(n-1)!}(1-e^{-\beta})^n e^{-\beta m}.
\label{pst}
\ee
Of course this formula can be used only for $n\geq 1$.

To analyze this result we first compute the \emph{conditional mean} number of quanta emitted given that $n$ are incident:
\be
\langle m|n\rangle_{\rm st}=\frac{(1-e^{-\beta})^{n}}{(n-1)!} \sum_{m=0}^\infty \frac{(m+n)!}{m!}\, e^{-\beta (m+1)}
\label{sum3}
\ee
Using \Eq{sum3'} of the Appendix we have
\be
\langle m|n\rangle_{\rm st}=\frac{n}{e^\beta-1}=n\langle m\rangle_{\rm sp}.
\ee
This is entirely analogous to the result in atomic physics that the mean number of quanta from stimulated emission is proportional to the number of incident quanta, with the proportionality coefficient equal to the mean number from spontaneous emission.  Our result here thus strengthens the claim that a black hole is capable of stimulated emission even for non-superradiant modes.

Actually in atomic physics one computes the \emph{probability} of stimulated emission of \emph{one} photon given that $n$ are incident, and finds it to be $n$ times the probability of one-photon spontaneous emission.   Were we to do likewise here we would get
\be
\frac{p_{\rm st}(1|n)}{p(1|0)}=n(1-e^{-\beta})^{n-1}.
\ee
This corresponds to the atomic physics result only for $n=1$ or when $\beta\to\infty$ for any $n\geq 2$.  Now by \Eq{spon} $\beta\to\infty$ is equivalent to $x\to\infty$, i.e., to the case $\hbar\omega\gg T$ (Wien regime).  In the intermediate and the Rayleigh-Jeans regimes the coefficient is smaller than that in the atomic physics calculation.  We interpret this discrepancy to reflect the fact that the mentioned atomic physics result is a first-order perturbation one, whereas \Eq{pst} here is nonperturbative.  This last is obvious since the formula in \Eq{spon} works for multi-quanta emission which would require application of arbitrarily high order perturbation theory.

\section{Composition of emission with scattering}
\label{sec:composition}

Let us now compose the distribution $p_{\rm st}(m|n)$ with that for spontaneous emission, $p(m|0)$, to get the conditional probability distribution for generic emission, $p_{\rm em}(m|n)$:
\be
p_{\rm em}(m|n)=\sum_{k=0}^m p_{\rm st}(k|n)\, p(m-k|0),
\ee
which equals
\be
(1-e^{-\beta})^{n+1}e^{-\beta m}\sum_{k=0}^m \frac{(n+k-1)!}{k!(n-1)!}.
\label{sum4}
\ee
The sum is worked out in the Appendix (\Eq{second}); we find
\be
p_{\rm em}(m|n)=\frac{(m+n)!}{m!n!}(1-e^{-\beta})^{n+1} e^{-\beta m}.
\ee

We note that $p_{\rm em}(m|n)=p_{\rm st}(m|n+1)$.  This accords with what happens in atomic physics: the total emission in the presence of $n$ quanta is as if $n+1$ quanta were responsible for purely stimulated emission, the extra unity being ascribed to the effect of vacuum fluctuations. 

We now compose the distributions $p_{\rm sc}$ and $p_{\rm em}$; this should give us $p(m|n)$:
\be
p(m|n)=\sum_{k=0}^{{\rm min}(m,n)}  p_{\rm em}(m-k|n)\,p_{\rm sc}(k|n),
\ee
where the upper limit of the summation reflects the fact that no more quanta can be scattered than the number incident but also cannot exceed the total number of outgoing quanta.  Thus
\be
p(m|n)=(1-e^{-\beta})e^{-(\mu n+\beta m)}\sum_{k=0}^{{\rm min}(m,n)}  \frac{(m+n-k)!(X-1)^{k}}{k!(m-k)! (n-k)!}  
\label{pmn3}
\ee
The coefficient of the sum here is exactly equivalent to that in \Eq{pmn2}.  Thus by focusing on the scattering contribution we have here derived an alternative form of $p(m|n)$, identical to that found in refs.~\cite{BekMeis,PananWald}, and which is also known to be equivalent to the $p(m|n)$ deduced in Sec.~\ref{sec:statistics} of this paper (see Sec.~\ref{sec:comparison}). 

\section{Summary}

It is widely accepted that a thermally radiating system cannot disclose details about its internal dynamics due to the loss of coherence implicit in the thermalization.  A black hole may be different in that, unlike ordinary matter, it is not demonstrably a collection of independent systems.   In this paper we have leaned on the hypothesis that a Kerr black hole's horizon area has a discrete spectrum to derive the black hole's response to incident radiation, specifically the conditional probability distribution for the number of emitted quanta.  We reproduce in two different ways the known probability distribution, obtaining to boot a closed form of it in terms of an hypergeometric function.  This accomplishment of the ``atomic'' model of the black hole shows that the widely hypothesized discreteness of the area spectrum is consistent with the expected statistics of the black hole radiance.   

Our approach here is related to the approach to quantum gravity of a black hole based on its quasinormal modes~\cite{BekMG,Hod,Maggiore}.   Quasinormal modes, however, are purely classical responses of the black hole to perturbations, and one cannot immediately discount the possibility that they are irrelevant to questions of quantum structure.  Nevertheless, the quasinormal mode approach does point to a discrete horizon area spectrum, often a uniformly spaced one~\cite{Skakala},~ and so agrees with our conclusion here that the black hole response to external radiation is consistent with a discrete area spectrum.

\appendix

\section{Sums and identities}
\label{sec:app}

Here we fill in various mathematical details needed in the paper.  

The inner sum in \Eq{inter} is done by expanding   
$(1-e^{-\beta})^{-1}$ in a series in $e^{-\beta}$, and then differentiating both sides $k$ times. We get
\be
\sum_{m=k}^\infty \frac{m!}{(m-k)!}e^{-\beta m}=\frac{k! e^{-\beta k}}{(1-e^{-\beta})^{k+1}}.
\label{first}
\ee

We now establish the equality of the two hypergeometric functions used in \Eqs{pmn1} and (\ref{pmn2}).  First we note that the function ${}_2F_1(a,b,c,z)$ solves the hypergeometric equation
\be
z(1-z)w''+[c-(a+b+1)z]w'-abw=0.
\ee
We note that if we replace $z\Rightarrow 1-z$ the equation 
goes over into the same equation, but with $c\Rightarrow -c+a+b+1$.
Hence ${}_2F_1(-m,-n,1,z)$ and ${}_2F_1(-m,-n,-m-n,1-z)$ are two solutions of the \emph{same} equation.  We now show they are the same solution up to normalization.  For this we use Eq.~15.3.6 of ref.~\cite{AbraSteg}:
\be
{}_2F_1(a,b,c,z)=\frac{\Gamma(c)\Gamma(c-a-b)}{\Gamma(c-a)\Gamma(c-b)}\ {}_2F_1(a,b,a+b+1-c,1-z)+\cdots
\ee
The omitted term has $\Gamma(a)$ and $\Gamma(b)$ in the denominator.  When we set $a=-m,b=-n$ and $c=1$ those $\Gamma$s blow up since $\Gamma(z)$ has poles at all non-positive integers; thus that term vanishes.
What is left is the identity we sought.

We now compute the sum in \Eq{sum3}.  In terms of the \emph{negative} variable $z=1-e^{\beta}$ it is is equivalent to
 \be
S\equiv (1-z)^{n}\frac{d^{n}}{dz^{n}} \sum_{m=0}^\infty \frac{1}{(1-z)^{m+1}} 
\ee
We first do the geometric sum,
\be
\sum_{m=0}^\infty \frac{1}{(1-z)^{m+1}}=(1-z)^{-1}[1-(1-z)^{-1}]^{-1}=-\frac{1}{z},
\ee
and then differentiate and replace the variable $z$ by $e^\beta$:
\be
S=n!\frac{(1-z)^{n}}{(-z)^{n+1}}=n!\frac{e^{\beta n}}{(e^\beta-1)^{n+1}}.
\label{sum3'}
\ee

Next we show that the sum in \Eq{sum4} is
\be
\sum_{k=0}^m\frac{(n+k-1)!}{(n-1)!k!}=\frac{(n+m)!}{n!m!}.
\label{second}
\ee
First note that for $m=0$ right and left sides of this equation equal unity.  Next assume the above equation is valid for some $m\geq 1$.  We now show that it is then valid for the next higher $m$.  If we raise the upper limit of the sum on the l.h.s. by unity, we evidently get for the full sum
\be
\frac{(n+m)!}{(n-1)!(m+1)!}+\frac{(n+m)!}{n!m!}=\frac{(n+m+1)!}{n!(m+1)!}.
\ee
But this equals the r.h.s. of \Eq{second} with $m\Rightarrow m+1$.  Hence, by induction, \Eq{second} is established for all $m$ and $n$.

\section*{Acknowledgments}

I thank Idan Talshir for a suggestive remark and Marcelo Schiffer for comments.  This research was supported by the I-CORE Program of the Planning and Budgeting Committee and the Israel Science Foundation (grant No. 1937/12), as well as by personal grant No. 24/12 of the Israel Science Foundation.


\begin{thebibliography}{99}

\bibitem{Hawking0}S.~W.~Hawking, Commun. Math Phys. D\textbf{43}, 199 (1975).

\bibitem{Parker}L.~Parker, Phys. Rev. D\textbf{12}, 1519 (1975).

\bibitem{Wald}R.~M.~Wald, Commun. Math Phys. D\textbf{45}, 9 (1975).

\bibitem{Hawking}S.~W.~Hawking, Phys. Rev. D\textbf{13}, 191 (1976).

\bibitem{BekMG}J.~D.~Bekenstein, in \emph{Proceedings of the Eight Marcel Grossman Meeting\/}, T. Piran and R. Ruffini, eds. (World Scientific, Singapore 1999), p. 92-111.

\bibitem{Corda}C. Corda, Eur. Phys. J. \textbf{73}, 2665 (2013).

\bibitem{spectrum}J.~D.~Bekenstein, Lett. Nuovo Cimento \textbf{11}, 467 (1974).

\bibitem{Mukh}V.~Mukhanov, Pis. Eksp. Teor. Fiz. \textbf{44}, 50 (1986) [Jour. Exp. Theor. Phys. Letters \textbf{44}, 63 (1986)].

\bibitem{BekMukh}
J.~D.~Bekenstein and V.~F.~Mukhanov, Phys. Letters B\textbf{360}, 7 (1995).

\bibitem{Maggiore}M. Maggiore, Phys. Rev. Letters \textbf{100}, 141301 (2008).

 \bibitem{Hod}S. Hod,  Phys. Rev. Letters \textbf{81}, 4293 (1998).

\bibitem{Fulfill}J.~D.~Bekenstein, in \emph{To Fulfill a Vision\/}, Y. Ne'eman, ed. (Addison-Wesley, Reading, Ma. 1981), pp.~42-59.

\bibitem{BekSchiff}J.~D.~Bekenstein and M.~Schiffer, Phys. Rev. Letters \textbf{72}, 2512 (1994).

\bibitem{BekMeis}J.~D.~Bekenstein and A.~Meisels, Phys. Rev. D\textbf{15}, 2775 (1977).

 \bibitem{AbraSteg}A.~Abramowitz and I.~A.~Stegun, \emph{Handbook of Mathematical Functions\/}, (Dover, New York  1970), pp.~559-561.

\bibitem{PananWald}P.~Panangaden and R.~M.~Wald, Phys. Rev. D\textbf{16}, 929 (1977).

 \bibitem{LLSP}L.~D.~Landau and E.~M.~Lifshitz, \emph{Statistical Physics\/}, Part. I (Pergamon, Oxford  1980), p.~161.

\bibitem{Skakala}J. Skakala, JHEP \textbf{1206}, 094 (2012).

\end{thebibliography}
\end{document}